\documentclass[fleqn,10pt]{wlscirep}
\usepackage[utf8]{inputenc}
\usepackage[T1]{fontenc}
\usepackage{lineno}
\usepackage{multirow}
\usepackage{gensymb}
\usepackage{makecell}
\usepackage{kotex}
\usepackage{kotex}

\title{DOO-RE: A dataset of ambient sensors in a meeting room for activity recognition}

\author[1]{Hyunju Kim}
\author[1]{Geon Kim}
\author[1]{Taehoon Lee}
\author[1]{Kisoo Kim}
\author[1,*]{Dongman Lee}
\affil[1]{Korea Advanced Institute of Science and Technology, School of Computing, Daejeon, 34141, South Korea}

\affil[*]{corresponding author(s): Dongman Lee (dlee@kaist.ac.kr)}

\begin{abstract}
With the advancement of IoT technology, recognizing user activities with machine learning methods is a promising way to provide various smart services to users. High-quality data with privacy protection is essential for deploying such services in the real world. Data streams from surrounding ambient sensors are well suited to the requirement. Existing ambient sensor datasets only support constrained private spaces and those for public spaces have yet to be explored despite growing interest in research on them. To meet this need, we build a dataset collected from a meeting room equipped with ambient sensors. The dataset, DOO-RE, includes data streams from various ambient sensor types such as Sound and Projector. Each sensor data stream is segmented into activity units and multiple annotators provide activity labels through a cross-validation annotation process to improve annotation quality. We finally obtain 9 types of activities. To our best knowledge, DOO-RE is the first dataset to support the recognition of both single and group activities in a real meeting room with reliable annotations.

\end{abstract}
\begin{document}

\flushbottom
\maketitle

\thispagestyle{empty}


\section*{Background \& Summary}


The proliferation of the Internet of things (IoT) enables surrounding spaces to support smart services such as public safety\cite{safety}, environmental monitoring\cite{monitoring} or independent living\cite{living}.
For this, intelligence used for understanding and recognizing users' intentions (i.e. human activity recognition) in a given space must be developed. 
Intelligence is formed from data streams generated in diverse sources such as cameras\cite{vision}, wearable sensors\cite{wearable}, or ambient sensors\cite{ambient} in the IoT domain.
To construct intelligence in a less labor-intensive manner, recent studies\cite{ar_methods, har_review} couple the data streams with machine learning techniques.
According to them, the importance of privacy-protected and high-quality data increases to improve the effectiveness of the techniques. 
User-identifiable features may be included while collecting data from given spaces, regardless of the intentions of users.
The user-identifiable information poses privacy invasions.
For example, the data streams from cameras contain users' heights or shapes\cite{camera_privacy} even after blurring and similarly, wearable devices could possibly include personal information such as heart rate or application usage patterns\cite{wearable_privacy}.
Data stream collection from surrounding ambient sensors is more appropriate for generic and real-world deployable data collection without breaching privacy.

The datasets based on the ambient sensors are listed in Table\ref{tab:relateddataset}.
The table summarizes the types of sensors, targeted activities, and types of users of each dataset (no user identification).
The related datasets (i.e. datasets except `DOO-RE') target private spaces such as smart homes with a small number of users and are recorded each user's actions respectively.
They are mainly constructed to focus on recognizing the simple activities of a single user.
Contrary to private spaces, public spaces\cite{public_space, smart_city} consist of more diverse and many users, and new types of activities such as group activities occur.
The situation of multiple users in a public space leads to more complex problems than in a single user-based space, which has inspired a lot of research\cite{multi-user, multi-user-2}. 

As research interest in public spaces increases, smart offices that help people perform their tasks efficiently and effectively also become one of the attractive research areas \cite{smart-office, smart-office-2}.
For example, if a user has to give a presentation on an important project, he or she only needs to focus on the main task, `presentation', and a smart service supports minor tasks such as `turning on a projector' or `turning off lights'.
To the best of our authors' knowledge, existing datasets\cite{energy, energy-2, energy-3} supporting office intelligence  focus only on observing energy consumption for entire workspaces while a dataset for recognizing users' intentions and building user-centric office intelligence is absent.
To meet the needs of a novel office intelligence dataset, we collect ambient sensor data streams from a meeting room, a type of smart office where both single-user and multi-user-based activities are conducted.

We release the DOO-RE dataset 
that consists of data streams collected 24/7 from a real-world meeting room, using a diverse set of sensors.
Unlike a home, a meeting room has different characteristics in terms of the physical structure (i.e. no purpose-wise physical divisions), the number of users (i.e. usually larger), and the duration of activities (i.e. longer in general).
The DOO-RE dataset is designed to capture sensor stream data incorporating such characteristics. For example, due to difficulty in discerning one activity from others with only location information, DOO-RE includes extra information on when and how long devices in a meeting room are used, whether there happens a change in the number of users, etc. For this, we employ a diverse set of sensors that can sense environmental changes, device activation, and user actions.
A collected sensor data stream representing one user activity is extracted as one activity episode in the DOO-RE dataset.
For trustworthy annotation, each episode's activity labels are annotated and validated by cross-checking and the consent of multiple annotators.
As a result, DOO-RE provides reliably labeled episodes for single and group activities from the meeting room.
DOO-RE is a novel dataset created in a public space. It contains the properties of a real-world meeting room. We consider that DOO-RE has a high potential to be good uses for developing powerful activity recognition approaches. 




\begin{table}[]
\centering
\begin{tabular}{|l|l|l|l|l|}
\hline
\textbf{Dataset name (years)} & \textbf{Ambient sensors} & \textbf{Target activities}  & \textbf{Types of users} \\
\hline
Wireless sensors & 14 wireless network nodes &  Leave house, Toileting, Showering, & Single user  \\
Dataset (2008)\cite{vankasteren} & & Sleeping, Preparing breakfast, etc.  &  \\
\hline
CASAS  & Motion, Item sensor, Temperature, & Read a magazine, Hang up clothes,  &  Two separate \\
(2009)\cite{casas} & Cabinet sensor, Burner sensor & Play a game, Set dining room table, etc.  & users\\
\hline
OPPORTUNITY & Switch, 3D acceleration sensor, & Groom, Relax, Prepare coffee,  &  Single user \\
(2010)\cite{opportunity} & 12 objects with 3D acceleration & Drink coffee, Prepare sandwich, etc.  & \\
\hline
ARAS & Force sensor, Distance, Photocell,  & Going Out, Preparing Breakfast, & Two separate\\
(2013)\cite{ARAS} &  IR, Contact sensor, Temperature & Having Breakfast, Preparing Lunch, etc.& users\\
\hline
ContextAct@A4H & Door, Light, Temperature, & Take Shower, Toilet use, Sleep, Cook, & Single user \\
(2017)\cite{context-act}&  Co2 levels, Appliances states & Leave Home, Wash Dishes, Eat, Work  & \\
\hline
E-care@home & Motion, Light, Pressure, & Sitting, Moving, Watching TV, Burning, & Single user \\
(2017)\cite{E-care} & Luminosity, Heartbeat simulator & Exercising, Cooking, Eating, etc. & \\
\hline
Motionless  & Accelerometer, Magnetometer, & Sleeping, Driving, Watching TV & Single user \\
Dataset (2022)\cite{motionless} &  Gyroscope, GPS, Microphone &  & \\
\hline
DOO-RE & Brightness, Humidity, Temperature,& Eating, Reading, Phone call, Seminar, & Single user,\\
(2023) &  Sound, Podium, Door, Motion, Seat,& Lab meeting, Small talk, Studying together, & Group user \\
 &  Aircon, Light, Projector & Technical discussion, Eating together &  \\
\hline
\end{tabular}
\caption{\label{tab:relateddataset} Comparison of the DOO-RE dataset and related ambient sensor-based datasets. 
The existing datasets focus on personal spaces, such as smart homes, which consist of a small number of users and where only simple activities occur.
Details of each dataset can be found in the corresponding citing paper.  }
\end{table}

\section*{Methods}

\subsection*{Dataset design}
As shown in Table \ref{tab:relateddataset}, the existing datasets except DOO-RE are designed to record the activity data generated by a single user or two users in a home, that is, a private space. 
They consist of data from only similar types of sensors (do not mix environment-driven and actuator-driven sensors). 
Applying these datasets to understand user behavior in a public space such as a meeting room may have limitations in terms of practical applicability and scalability.
That is a key reason for the necessity of a new dataset like DOO-RE.

To meet such a need, we construct a dataset collected from a meeting room equipped with ambient sensors. 
We name our dataset `DOO-RE', a traditional Korean word for `people cooperate', as a meeting room is considered a place for accommodating various group activities.
The dataset is useful for user behavior analysis with two objectives as follows:
\begin{enumerate}
  \item Provide a foundation for understanding how ambient sensors are excited in the presence of multiple users.
  \item Explore behavioral patterns of users in a real smart meeting room that help construct feasible activity recognition methods.
\end{enumerate}


A meeting room has different characteristics compared with a smart home which the related datasets focus on.
First, unlike a smart home that is usually \textit{physically} divided by multiple sub-regions with clear purposes of use (e.g. kitchen and bedroom), it is not easy to divide a meeting room into physically independent sub-regions by their purposes.
In other words, the former can discern users' activities with a specific type of user-driven sensor (i.e. location sensors) alone, while the latter cannot.
Therefore, the dataset should include data from other types of information sources that help figure out what activities are conducted by users.
For example, we can leverage device usage information from actuator-driven sensors such as light on/off or projector on/off sensors, and environmental state changes information from environmental-driven sensors such as sound and temperature sensors to distinguish `Seminar' activity from others. 
Secondly, multiple users perform various actions simultaneously in an activity.
To capture simultaneous actions performed by multiple users, it is necessary to record the duration information of user-driven sensors (e.g. seat sensor) or actuator-driven sensors (e.g. projector on/off sensor) related to each action.
The duration of each sensor is recorded in terms of the start and end states of each sensor unlike the previous datasets, which only record the start states of each sensor. 
For example, if the `Seat \#1' sensor starts at timestamp 1 and ends at timestamp 900, and the `Seat \#2' sensor starts at  timestamp 100 and ends at timestamp 1000, we can detect that these are simultaneous actions.
Thirdly, members can change in the middle of an activity, which causes great variation between the episodes of the same type of activity. 
This requires multiple types of user-driven sensors - location sensors as well as seat sensors -
that indirectly describe member changes without compromising privacy.
We record data from sensors that detect user state changes such as `a user being seated' or `a user in and out' in the meeting room. 
Finally, activity duration in a meeting room is usually longer than in a home since the number of participants is higher.
This implies that an efficient data structure should be designed not to incur storage shortages.
Thus, we record data from user-driven and actuator-driven sensors only when their states change rather than save all raw sensor data.
To capture the above characteristics of a meeting room, we design our dataset to record various types of sensors such as environment-driven, user-driven, and actuator-driven.

DOO-RE is an annotated dataset that consists of sensor data obtained in the wild, necessitating the label selection and validation steps. 
It is different from the existing ambient sensor-based datasets assuming that users only perform activities in pre-determined categories.
In other words, relying on one expert to name or annotate activities is risky in this environment because names and views for the same activity can vary from person to person.
The solution is to ensure that a majority of people agree on the name and view of the activity and that the annotation results fit accordingly.
We select multiple annotators and they harmonize terms for each activity label through a discussion process.
Then, they annotate unnamed episodes based on the activity labels and validate the labeling results together.

\subsection*{Data collection setup}

\subsubsection*{A meeting room condition}
\begin{figure}[h]
    \centering
    \includegraphics[width=0.9\textwidth]{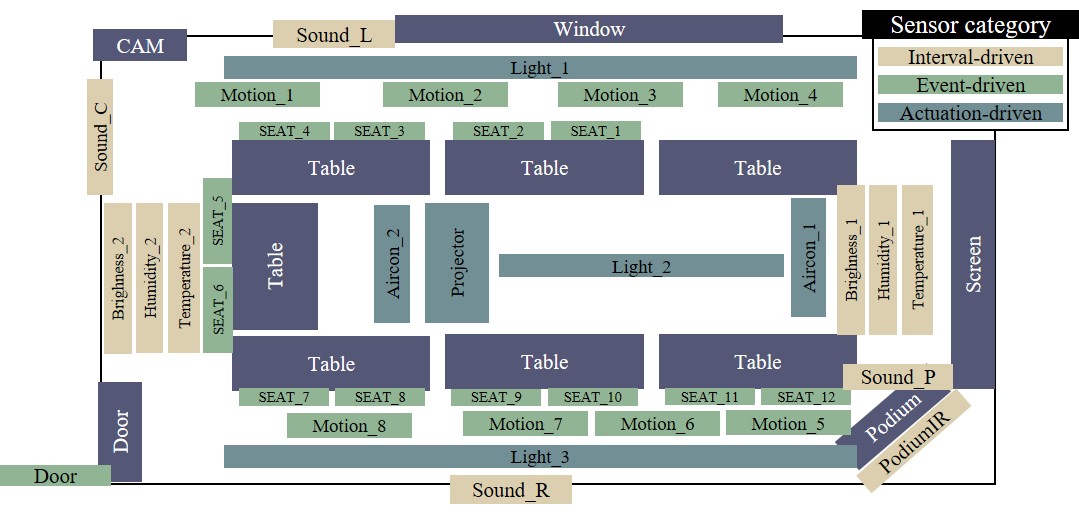}
    \caption{The meeting room testbed layout and installed location of ambient sensors.
    Stuff related to meetings, such as tables and a projector, are basically arranged.
    Various types of ambient sensors are installed to track user actions.
    We attach suitable types of sensors for each area after observing user activities in the meeting room.}
    \label{fig: sensor_location}
\end{figure}

Ambient sensors, which generate data based on users' behavior sequences, are installed in a university meeting room.
As shown in Figure \ref{fig: sensor_location}, the meeting room is a typical meeting space with one door, seven long tables, and two seats per table. 
A projector, a screen, and a podium are provided to help users give a presentation.
We set the left or right direction, depending on where a user is looking at the screen. 
The wall with a window is called the left wall.
For example, `Sound\_L' refers to a sound sensor installed on the left wall.

Graduate students and faculty members primarily use the meeting room.
There are no specific criteria such as age, race, or gender for entry and participation in activities.
Unlike the existing datasets, we do not give any action guidelines to capture realistic user behavior patterns in the space.
Users naturally perform their actions according to their roles in a participating activity.
As a result, we have a diverse set of activities conducted by single or multiple users whose ages vary from 20s to 60s.
We believe that this enables DOO-RE to have more generality and applicability than the existing ambient sensor-based datasets.

\subsubsection*{Installed ambient sensors description}

We employ various types of ambient sensors that help us understand what activities users perform. 
The sensors are classified into three categories: Environment-driven, User-driven, and Actuator-driven.
Environment-driven sensors refer to sensors that monitor environmental changes in the space.
User-driven sensors are a category of sensors that publish values when user states change.
Actuator-driven sensors detect whether the states of the corresponding actuators have changed.

We observe our meeting room testbed for figuring out how users exploit the meeting room for a month and find 9 different activities as shown in Table \ref{tab: activity information}.
Based on this, we draw a map for what types of ambient sensors should be installed as shown in Figure \ref{fig: sensor_location}. 
This setup is well configured to capture the sequences of sensor data when certain activities are performed.

We install four types of environment-driven sensors: Brightness, Humidity, Temperature, and Sound.
\textbf{Brightness}, \textbf{Humidity} and \textbf{Temperature} sensors are attached in each sub-region where internal environment information can be represented differently depending on changes in the external environment (e.g. changes in weather projected through the window).
For each type, two sensors are installed in front of and behind the meeting room and they give different values depending on the external environment.
\textbf{Sound} sensors are installed at fixed positions on each wall to track which side the target sound is coming from. 
They are also attached to certain objects to capture specific user actions (e.g. presentation).

Three types of user-driven sensors exist in the meeting room: Presenter Detection, In/Out, Motion, and Seat.
\textbf{Presenter Detection} sensor is stuck to the front of the podium to recognize a presenter.
\textbf{In / Out} sensor is located at the entrance of the meeting room to detect the user entering or exiting.
\textbf{Motion} sensors are attached to the walls along a user's movement path in the meeting room.
\textbf{Seat} sensors are placed on each seat to detect if the seat is occupied by a user. 

Actuator-driven sensors are attached to their corresponding devices to determine whether a device is in use. \textbf{Aircon} sensors are attached to each of the two air conditioners on the ceiling of the meeting room. For the \textbf{Light} sensor, a light switch is on the right side of the entrance, and the three light groups are located on the ceiling of the meeting room. For \textbf{Projector} sensor, the control panel of the projector is on the podium, and the projector is in the center of the ceiling.
  
If multiple sensors of the same type are installed in one space, different sensor names should be given to distinguish them as shown in \textbf{Sensor Name} of Table \ref{tab:category-sensors}.
We distinguish them by adding numbers or alphabets and apply the finally determined sensor names to the dataset record file.
Each sensor's name is closely related to the sensor's location.

\begin{table}[t]
\centering
\begin{tabular}{|c|c|c|}
\hline
\textbf{Sensor Categorization} & \textbf{Sensor Type} & \textbf{Sensor Name}\\ \hline
\multirow{4}{*}{Environment-driven} & Brightness & Brightness\_1, Brightness\_2 \\ \cline{2-3}
                                    & Humidity & Humidity\_1, Humidity\_2\\ \cline{2-3} 
                                    & Temperature & Temperature\_1, Temperature\_2 \\ \cline{2-3} 
                                    & Sound & Sound\_L, Sound\_C, Sound\_R, Sound\_P \\ \hline
                                    
\multirow{6}{*}{User-driven}        & Presenter Detection  & PodiumIR \\ \cline{2-3}
                                    & In / Out & Door \\ \cline{2-3} 
                                    & Motion & \makecell{ Motion\_1, Motion\_2, Motion\_3, Motion\_4, \\ Motion\_5, Motion\_6, Motion\_7, Motion\_8}  \\ \cline{2-3}     & Seat & \makecell{Seat\_1, Seat\_2, Seat\_3, Seat\_4, Seat\_5, Seat\_6,\\ Seat\_7, Seat\_8, Seat\_9, Seat\_10, Seat\_11, Seat\_12} \\ \hline
                                    
\multirow{3}{*}{Actuator-driven}    & AirCon & Aircon\_1, Aircon\_2 \\ \cline{2-3}
                                    & Light & Light\_1, Light\_2, Light\_3 \\ \cline{2-3}
                                    & Projector & Projector \\ \cline{2-3} \hline                      

\end{tabular}
\caption{Categorization of ambient sensors installed in the meeting room. 
        \textbf{Environment-driven sensors} are sensors that track changes in environmental conditions.
        \textbf{User-driven sensors} are sensors that capture changes in users' states.
        \textbf{Actuator-driven sensors} are sensors that display status when users operate actuators.
        }
\label{tab:category-sensors}
\end{table}

\begin{figure}[h]
    \centering
    \includegraphics[width=0.9\textwidth]{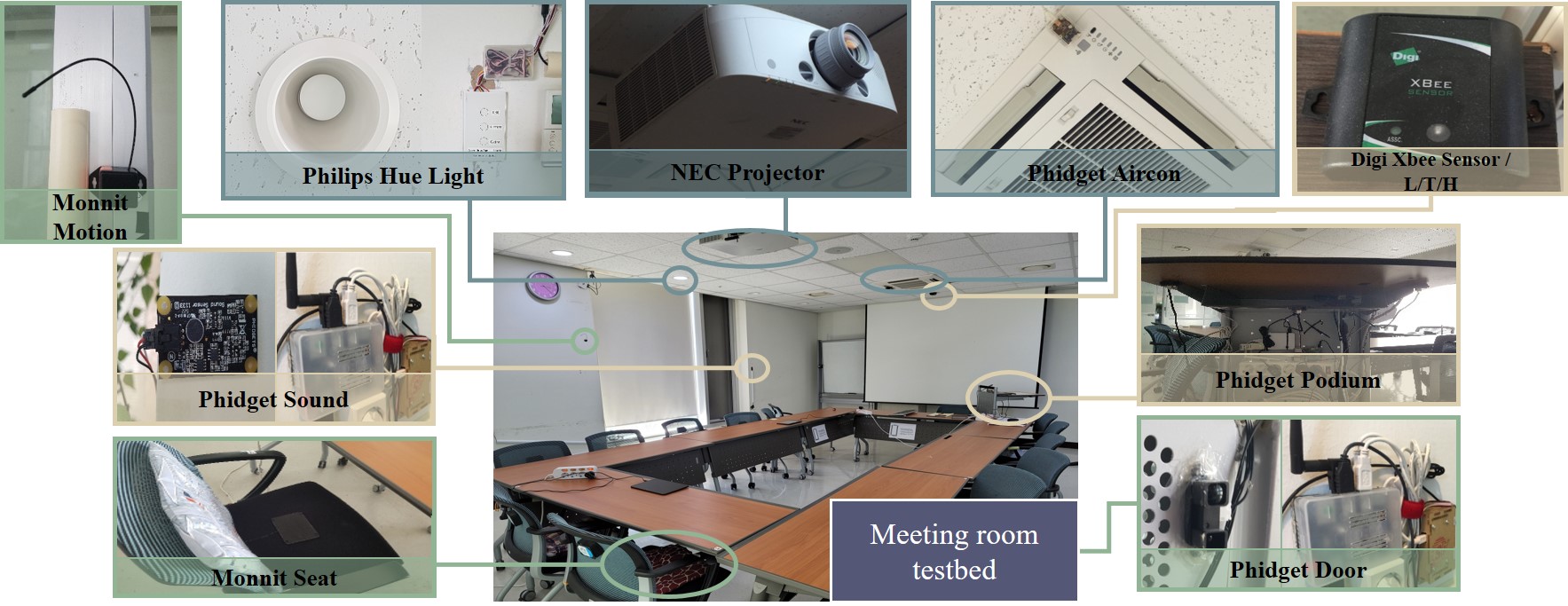}
    \caption{Sensors located in the meeting room testbed.
    \textit{Digi XBee Sensor /L/T/H} and \textit{Phidget} are used for environmental sensors.
    \textit{Phidget} and \textit{Monnit} are leveraged for user-driven sensors.
    \textit{Phidget} and \textit{API} are utilized for actuator-driven sensors. }
    \label{fig: physical_device}
\end{figure}

\begin{table}[h]
\centering
\begin{tabular}{|c|c|c|c|}
\hline
\textbf{Product Name} & \textbf{Sensor Type} & \textbf{Publish Time} & \textbf{Value Range} \\ \hline
\multirow{3}{*}{Digi XBee Sensor /L/T/H} & Brightness & 60s & 0 - 1100 \\ \cline{2-4} 
                                         & Humidity & 60s & 0 - 100\%\\ \cline{2-4} 
                                         & Temperature & 60s & 15 - 30 \degree C \\ \hline
\multirow{4}{*}{Phidget} & Sound & 10s &  34 - 102 dB  \\ \cline{2-4}
                         & Presenter Detection & 10s & 25 - 400 \\ \cline{2-4} 
                         & In / Out & State Changes  & Activate\\ \cline{2-4}                   
                         & Aircon & State Changes & ON / OFF \\ \hline
\multirow{2}{*}{Monnit} & Motion & State Changes & True / False \\ \cline{2-4}                       
                         & Seat & State Changes & True / False \\ \hline
            {Philips Hue} & Light & State Changes & ON / OFF \\ \cline{2-4}\hline                       
            {NEC Projector} & Projector & State Changes & ON / OFF \\ \cline{2-4} \hline                         

\end{tabular}
\caption{Summary of product name, sensor type, publish time, and value range of each sensor. 
Publish times and value ranges vary according to the sensor types. 
\textbf{Environment-driven} and \textbf{Presenter Detection sensors} publish their states at regular intervals and their ranges of values are varied. 
On the other hand, \textbf{user-driven (excluding Presenter Detection)} and \textbf{actuator-driven sensors} have only two states as values and publish their states only when the values are flipped. }
\label{tab:physical-devices}
\end{table}

Figure \ref{fig: physical_device} shows what kinds of sensor products we use in our meeting room testbed for implementing the various types of sensors mentioned above. 
We use both wired and wireless sensors. 
For the ease and stability of data collection, wireless sensors are applied as a priority.
Wired sensors are used only when certain types of wireless sensors are not available or there are practical issues such as battery drain. The following are the details of the sensor products we employ:

\begin{enumerate}
  \item 
    \textit{Digi XBee Sensor /L/T/H} is a wireless integrated ambient light, temperature, and   humidity sensor.
    It derives numerical values of \textbf{Brightness}, \textbf{Temperature}, and \textbf{Humidity} near the sensor.
    It transmits data to a Digi XBee Gateway via XBee network infrastructure.
    We set the sensor status publishing time to every 60 seconds in order to avoid battery drainage.
  
  \item 
    \textit{Phidget}s are wired devices for sensing sound or IR and work on top of Raspberry Pi.
    IR sensors are installed at three places: podium, door, and air-conditioner. 
    Their purposes are to perform the role of \textbf{Presenter Detection}, \textbf{In / Out}, and \textbf{AirCon}, respectively as described in Table \ref{tab:category-sensors}.
    \textbf{Sound} sensor generates numerical sound pressure values (dB) and each sound sensor publishes its status every 10 seconds for reliable data collection. 
    \textbf{Presenter Detection} sensor uses reflected IR light to measure the distance of a certain object from a podium. 
    Even if it is a user-driven sensor, it generates a sensor state every 10 seconds because the distance between the podium and the presenter can be important. 
    \textbf{In / Out} sensor also employs reflected IR light to check user access to the space. 
    Whenever a user passes through a door, the sensor publishes that people have entered or left the meeting room.
    \textbf{In / Out} sensor publishes data when its state changes.
    The sensor may help determine when an activity starts and ends and estimate the number of people involved in each episode of activities.    
    \textbf{Aircon} sensor reads the power state of an air conditioner using Phidget IR controllers.   
    It publishes the sensor value depending on the status of the air conditioners.
    
  \item
    \textit{Monnit} wireless sensors realize motion or seat occupation sensing. 
    \textbf{Motion} sensor detects user movement through PIR technology and generates its value by the sensor state changes. 
    Since the sensors have 80-degree viewing angles, we stick them to the wall without interfering with each other.
    \textbf{Seat} sensor captures seat occupancy by using pressure imparted to a plate. 
    Its value is derived from pressure value into a boolean value;  that is, true if a person is seated on its corresponding target chair; otherwise, false.

  \item \textbf{Light} sensor retrieves the power states of three light groups (Left, Center, Right) via the \textit{Philips Hue} API. 
 \textbf{Projector} sensor reads the power state using with the \textit{NEC Projector} API.
  The API provides an interface to figure out if the user has changed the state of the projector.
\end{enumerate}

A summary of product name, sensor type, publish time, and value range for each sensor type is described in Table \ref{tab:physical-devices}. 
To transmit and store the generated sensor data in a structured form (i.e., tuple) to 
the database, we implement a software agent that runs on a Raspberry Pi or mini PC networked with sensors.
A tuple of the sensor type, sensor name, sensor value, and data-generated timestamp is sent as a data point.

\subsection*{Ethics statement}
Collecting sensor data of the DOO-RE dataset is approved via the Korea Advanced Institute of Science and Technology (KAIST) Institutional Review Board (IRB).
The IRB investigates the purpose of the data collection, the subjects, the collection method, the observed testbed environment, the types of data collected, the safety evaluation and solutions for side effects, etc.

During the data collection time, we notify in advance users that we collect data in the meeting room.
Note that data is deleted from the database when a user does not consent.
When storing ambient sensor data, user-identifiable information is not saved in the database.
Video data for annotations are stored in the database after blur processing, making them impossible to identify users.
The video data are deleted from the database subsequent to annotating.

\subsection*{Data annotation procedure}
Three annotators who are experts in the meeting room annotate activity labels by six steps after data collection from the installed ambient sensors: 1) Data refinement, 2) Segmentation, 3) Activity label selection, 4) Activity annotation, and 5) Annotated label validation.

\subsubsection*{Data refinement}
For efficient processing of later steps, we fetch sensor data from the database and leave only data points that are meaningful for annotation.
In the database, each sensor data point is stored in the form of a tuple \{timestamp, sensor\_type, sensor\_name, sensor\_value\}, and videos with user information obscured are also kept for annotation.
We extract all ambient sensor and video data from the first 4 months of IRB approval. 
Sensor data points without video are eliminated as they cannot be clearly annotated.
Data points produced at periods when particular activities do not occur are removed since they are not subject to annotation.
The human tracking method \cite{object_detection} helps to roughly determine, based on the recorded video, whether a certain activity takes place in the meeting room.
This step ultimately leaves only sequences of data points worth annotating.

\subsubsection*{Segmentation}
To clarify targets for data annotation, a segmentation step divides the refined data sequences to define the start and end times of each episode.
An episode is a one-sample sensor data sequence for a certain activity.
First, for labor efficiency, we run a program that performs an approximate segmentation. 
Within incoming data sequences, the program automatically detects the timestamps with large deviations from certain sensors' values, such as `In / Out' or `Sound' sensors. 
Human Annotators begin to take part from this step and we distribute the segmentation task according to their numbers.
They watch the video recorded at those timestamps to find out the exact start and end times of each episode.
As a result, episodes not yet named are extracted and the start times and end times of each episode are recorded in a spreadsheet.
After the annotators' segmentation, we remove episodes that are shorter than 5 minutes, the minimum amount of time users can engage in a meaningful activity.
During the segmentation, for use in the activity label selection step, each human annotator writes down a list of occurring activities and a brief description of them. 

\subsubsection*{Activity label selection}
To ensure terminological consistency of activity names among them, annotators discuss activities seen in the segmentation step and bundle the same-typed activities with the same terms to produce final activity labels.
First, each annotator shares the activities in their lists of activities recorded in the segmentation step.
This procedure is necessary not only to navigate through all activities taking place in the meeting room but also to adjust the view equally among annotators on the same activity.
Annotators classify activities by the user behavior sequence similarity and discuss which terms are appropriate for each activity (i.e., labels).
Some activities in exceptional cases that rarely appear in the meeting room are discarded.
As a result, we finalize 9 activity labels to be annotated: 3 for single-user based and 6 for group-user based as shown in Table \ref{tab: activity information}.
The characteristic of each activity is described as follows:
\begin{itemize}
     \item Single-user-based activities
     \begin{itemize}
      \item \textbf{Eating}: a single person comes in and eats quietly at one table.
      \item \textbf{Phone call}: a single person walks around the meeting room, stands still or sits, and makes a phone call.
      \item \textbf{Reading}: a single person sits at a table to read or study.
     \end{itemize}
     \item Group-user-based activities
     \begin{itemize}
      \item \textbf{Small talk}: a pair of people sit down nearby and have a casual conversation.
      \item \textbf{Studying together}: two or more people study together for a long time.
      \item \textbf{Eating together}: three or more people come in and eat together at multiple tables.
      \item \textbf{Lab meeting}: a regular meeting where the majority of attendees take turns presenting.
      \item \textbf{Seminar}: a meeting in which an invited presenter(s) presents a specific topic.
      \item \textbf{Technical discussion}: a meeting in which a group of people discusses technical issues.      
     \end{itemize}
\end{itemize}

\subsubsection*{Activity annotation}
In the activity annotation step, the annotators assign activity labels to the segmented episodes.
We divide the resulting number of episodes by the number of annotators and assign the same number of episodes to each annotator.
Each annotator becomes a primary annotator for the episodes assigned at this time.
The primary annotator views the video data and assigns an activity label to each episode.
For the efficiency of annotation, they quickly determine an activity type of each episode by dividing its video data into beginning, middle, and end parts, and figuring out a rough sequence of users' actions derived therefrom.
Annotation accuracy is also important, giving the primary annotators ample time to judge activity labels.
Finally, all episodes are assigned activity labels and the annotation results are recorded in the spreadsheet used in the segmentation step.
They also write down any points of discussion that arise when annotating.

\subsubsection*{Annotated label validation}
For activity labels annotated by each primary annotator, other annotators verify the suitability of those labels as secondary annotators.
Secondary annotators are assigned episodes that were not annotated in the previous step.
For each episode, each secondary annotator looks at random parts of its video and verifies that they agree with the activity label of the primary annotator.
They are encouraged to judge the labels within 5 minutes to avoid long pondering.
If secondary annotators have different opinions about labels for particular episodes, they write the opinions in the spreadsheet.
After the validation process for all episodes, if multiple annotators annotate an episode with the same activity label, this agreed label is assigned to that episode.
We confirm that about 90\% of the episodes correspond to this case.
In the case of an episode judged to be a different label even one of the annotators, a discussion session is conducted.
If an activity label is derived to be agreed upon between annotators, it is given the name of the episode, and if no label is found to reach an agreement, the episode is removed as it may affect the overall quality of the dataset.
Finally, when all episodes are reliably labeled, we generate a sensor data sequence file for each episode, referring to the start and end times obtained at the segmentation step.
This efficient and accurate data annotation procedure finally builds DOO-RE, an ambient sensor-based dataset.

\begin{table}[]
\centering
\begin{tabular}{|c|c|c|c|c|}
\hline
\textbf{Category} & \textbf{Activity name} & \textbf{\# of episode} & \textbf{Average duration (sec)} & \textbf{Average \# of participants} \\ \hline
\multirow{3}{*}{Single-user based}  & Eating & 42 & 1440.60 & 1 \\ \cline{2-5} 
                                    & Phone call & 150 & 807.15 & 1 \\ \cline{2-5} 
                                    & Reading & 213 & 2439.25 & 1 \\ \hline
\multirow{6}{*}{Group-user based}   & Small talk & 153 & 2786.79 & 3 \\ \cline{2-5} 
                                    & Studying together & 21 & 4925.05 & 2 \\ \cline{2-5} 
                                    & Eating together & 16 & 1482.44 & 3 \\ \cline{2-5} 
                                    & Lab meeting & 29 & 4515.62 & 10 \\ \cline{2-5} 
                                    & Seminar & 33 & 3332.82 & 5 \\ \cline{2-5} 
                                    & Technical discussion & 39 & 3379.67 & 7 \\ \hline
\end{tabular}
\caption{The activity annotation resulting after annotated label validations step. 
We find 9 activities in the meeting room: 3 for single-user and 6 for group-user based.
A \textbf{single-user-based} activity is for one person to perform a single task objective, and a \textbf{group-user-based} activity is for several people to collaborate on one task.
The number of episodes, average duration, and the average number of participants vary from activity to activity.
}
\label{tab: activity information}
\end{table}



 

\section*{Data Records}


\begin{table}[t]
\centering
\begin{tabular}{|c|l|}
\hline
\multicolumn{2}{|c|}{\textbf{A summary of data collection}} \\
\hline
Episode duration & Total: 1627406 sec (mean: 2338.22, median: 1443, stdev: 2379.93)\\
\hline
Number of episodes & Total: 696 (mean: 77.33, median: 39, stdev: 69.42)\\
\hline
Number of participants & Total: 1655 (mean: 2.38, median: 1, stdev: 2.73)\\
\hline
\multirow{3}{*}{Activity categories} & \textbf{Single-user based}: Eating, Phone call, Reading  \\\cline{2-2}
& \textbf{Group-user based}: Small talk, Studying together, Eating together,\\
& Lab meeting, Seminar, Technical discussion \\ 
\hline
\multirow{3}{*}{Categories of sensors} & \textbf{Environment-driven}: Brightness, Humidity, Temperature, Sound \\\cline{2-2}
& \textbf{User-driven}: Presenter Detection, In / Out, Motion, Seat \\\cline{2-2}
& \textbf{Actuator-driven}: Aircon,  Light, Projector\\ 
\hline
\end{tabular}
\caption{\label{tab:collected result}The summary of data collection results. 
To characterize the DOO-RE dataset, we calculate total, mean, median, and standard deviation values for episode duration, number of episodes, and number of participants, respectively.
Categories of activities and sensors appearing in DOO-RE are also organized for easy viewing at a glance.}
\end{table}

\subsection*{Overall data record description}
As shown in Table \ref{tab: activity information}, DOO-RE contains 9 types of activity data, each with its own number of episodes, average duration, and average participants.
The total episode time for the DOO-RE dataset's 696 episodes is 452 hours (1627406 seconds) and 1655 participants perform the activities (the same person may be counted in duplicate).
The meeting room consists of various sensor categories such as environment-driven, user-driven, and actuator-driven, and each activity is composed of consists of its own sensor combination according to its characteristics.
A summary of the characteristics of the collected dataset is summarized in Table \ref{tab:collected result}.
We provide a useful website (\url{http://doo-re.kaist.ac.kr/}) to our readers, and DOO-RE is available upon request from that site and the figshare repository (\url{https://figshare.com/s/52e9956527539cc2aa63}).

DOO-RE is based on KST (Korea Standard Time) Unix timestamp (UTC + 9:00) due to the location of the meeting room testbed (i.e., Republic of Korea).
In the other words, in order to obtain the precise time at which a particular activity or sensor occurred, its timestamp must be converted to a human date based on the KST time zone.
The timestamps are given in millisecond units to clearly distinguish between sensor data that is published almost simultaneously.

DOO-RE's directories and files are organized as follows:
DOO-RE has a root directory named `DOO-RE' and it consists of 9 types of activity directories named: \textit{Eating}, \textit{Phone call}, \textit{Reading}, \textit{Eating together}, \textit{Lab meeting}, \textit{Seminar}, \textit{Small talk}, \textit{Studying together}, \textit{Technical discussion}. 
Each activity directory consists of \textit{metadata} and \textit{sensor} sub-directories, each containing metadata files and sensor data files for each episode.

Metadata refers to overall information about an episode, including timestamps at the start and end of the episode, the duration of the episode, the average number of participants during the episode, and all sensor names that have appeared during the episode.
A sensor data file records the values of the sensor data points actually published during the episode. 
By formatting the file names of \textit{metadata} directory and \textit{sensor} directory as <Activity\_name>\_<index>.txt and <Activity\_name>\_<index>.csv, respectively,
the metadata files and sensor data files of each episode form a one-to-one correspondence based on the same index.
For example, `DOO-RE/Seminar/metadata/Seminar\_3.txt' is a metadata file, and `DOO-RE/Seminar/sensor/Seminar\_3.csv' is a sensor data file for the `Seminar\_3' episode.
It also implies that metadata files are in `txt' format and sensor data files are in `CSV' format.

\subsection*{Metadata}
Metadata files contain auxiliary information about each episode that cannot be captured in sensor data files.
It is possible to understand episodes' general traits by looking at the metadata.
A metadata file `<Activity\_name>\_<index>.txt' contains the following content:
\begin{enumerate}
    \item \textbf{label} is the annotated activity label of the episode. 
    \item \textbf{start\_ts (msec)} is timestamp of the episode start. The timestamp is relative to UTC +9:00.
    \item \textbf{end\_ts (msec)} is timestamp of the episode end. The timestamp is relative to UTC +9:00.
    \item \textbf{Avg\_n\_humans} is the average number of people participating in the activity. This value is measured during the activity annotation step. 
    \item \textbf{Duration (sec)} is duration of the episode. It is simply calculated as (end\_ts - start\_ts) / 1000.
    \item \textbf{Sensors} is a list of the sensor names that appear during the episode. 
\end{enumerate}

\subsection*{Sensor data}
A sensor data file has the data point sequences of one activity episode.
In a file, each data point consists of three columns: \textbf{timestamp}, \textbf{sensor name}, and \textbf{sensor value}.
The \textbf{timestamp} refers to the time at which a sensor value is generated, and the \textbf{sensor name} is a term that can be identified between the sensors based on the location of each sensor, which can be referred to Table \ref{tab:category-sensors}.
The location of each sensor corresponding to each \textbf{sensor name} is in Figure \ref{fig: sensor_location}.
The \textbf{sensor value} means a state of  a sensor at that timestamp and is within the value range shown in Table \ref{tab:physical-devices}.
As mentioned before, ambient sensors are categorized into three classes: Environment-driven, User-driven, and Actuator-driven.
To avoid running out of storage, if possible, we store data from User-driven and Actuator-driven sensors only when their state changes, rather than all raw data.

\subsubsection*{Environment-driven}
\begin{figure}[h]a
    \centering
    \includegraphics[width=\textwidth]{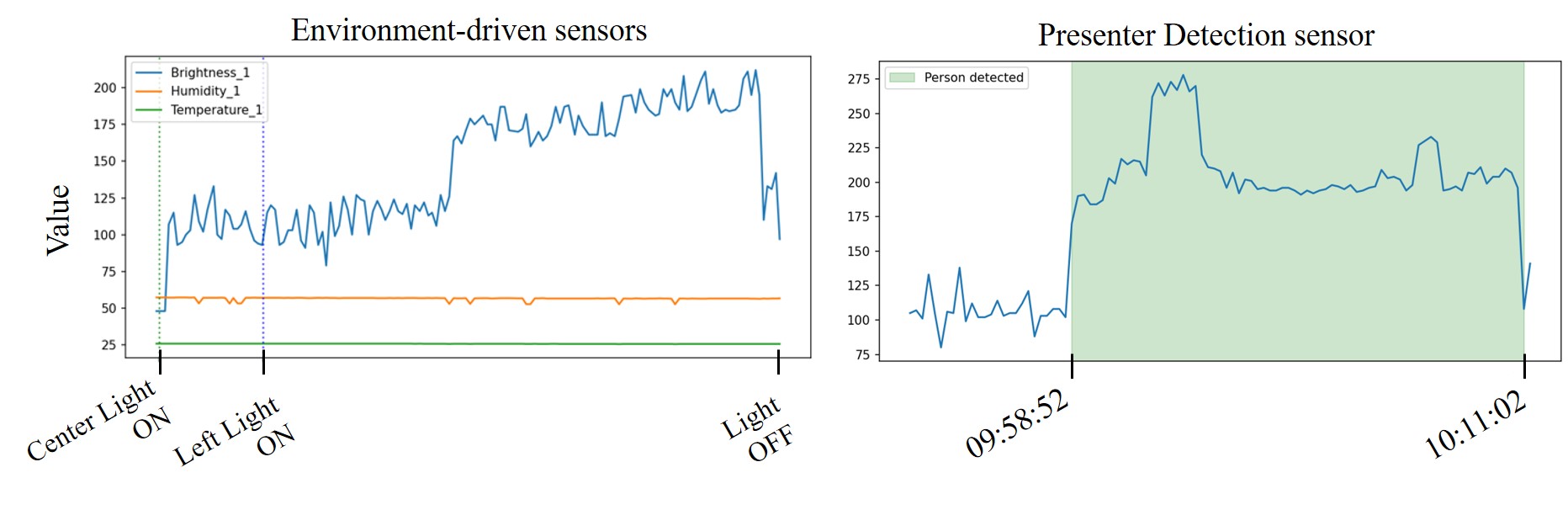}
    \caption{An illustration of a value change graph for environment-driven sensors (i.e. Brightness, Temperature, and Humidity sensors) and the `Presenter Detection' sensor in the file `Seminar\_0.csv'.
    The figure on the left shows how each \textbf{environment-driven sensor} state can change in one episode. The figure on the right shows that the value of the \textbf{`Presenter Detection' sensor} increases when there is a presenter in front of the podium.}
    \label{fig: sensor_podiumir}
\end{figure}

Each environment-driven sensor measures the environmental conditions in the meeting room at regular intervals. 
\textbf{Brightness}, \textbf{Humidity}, \textbf{Temperature} and \textbf{Sound} sensor types are the environment-driven in DOO-RE.
\begin{enumerate}
\item \textbf{Brightness}, \textbf{Humidity}, \textbf{Temperature} -
Each sensor type is installed on the front and back of the meeting room, appending `1' to the names of the front side sensors and `2' to the names of the back side sensors.
\textbf{Brightness\_1} and \textbf{Brightness\_2} measure brightness based on the light intensity around each sensor to determine whether each sub-region is bright or dark.
Their values range from 0 to 1100, with higher values meaning brighter.
\textbf{Temperature\_1} and \textbf{Temperature\_2} determine the temperature near each sensor in degrees Celsius (\degree C).
Higher values indicate greater heat and their values range from 15 to 30.
\textbf{Humidity\_1} and \textbf{Humidity\_2} calculate the relative humidity (\%RH) near each sensor.
These sensors have high values when the moisture content in the air is high, producing a value between 0 and 100. 
Figure \ref{fig: sensor_podiumir} draws an example of these sensors' value changes during an activity episode.
\newline
Their data points are represented as `Brightness\_X', `Temperature\_X', or `Humidity\_X' in the `sensor\_name' column of the CSV files, where X is 1 or 2. 
And values of each data point are recorded in the `value' column of the CSV files.

\item \textbf{Sound} -  
Each sound sensor measures the sound pressure level near each location.
The meeting room is equipped with four sensor sensors, each sensor name associated with its location.
\textbf{Sound\_P} is installed in front of a podium in the meeting room.
\textbf{Sound\_C} is installed on the south wall of the meeting room.
\textbf{Sound\_R} is installed on the right wall of the meeting room.
\textbf{Sound\_L} is installed on the left wall of the meeting room.
The sensor value has a higher value when there is a loud sound in the space.
\newline
All data points of sensors are indicated as `Sound\_X' in the `sensor\_name' column of the CSV files, where X is P, C, R, or L. 
Their values are displayed in the `value' column of the CSV files and range from 34 dB to 102 dB.
\end{enumerate}

\subsubsection*{Users-driven}
User-driven sensors detect user state changes in the meeting room. 
Four types of sensors are installed in the meeting room: \textbf{Presenter Detection}, \textbf{In / Out}, \textbf{Motion}, and \textbf{Seat}.

\begin{enumerate}
\item \textbf{Presenter Detection} - The sensor is located in front of the podium and measures the distance of objects up to 30 cm away from the podium.
We record the sensor's data points under the name \textbf{PodiumIR}.
Based on our empirical knowledge, if a person doesn't exist near the podium, its value keeps fluctuating between 50 and 150. 
If a person exists in front of the podium, the sensor generates larger values as shown in Figure \ref{fig: sensor_podiumir}.
\newline
All data points of the sensor are indicated as `Podium' in the `sensor\_name' column of the CSV files. 
Their values are between 25 and 400 and appear in the `value' column of the CSV files.

\item \textbf{In / Out} - 
The sensor refers to the access of users which is installed at the entrance of the meeting room and is named \textbf{Door} as the sensor name.
An `active' value is generated when it detects people near the door.
\newline
All data points are recorded as `Door' in the `sensor\_name' column and `active' in the `value' column in the CSV files.
\item \textbf{Motion} - Each motion sensor detects user movement near itself and is given a specific number in its name based on its location (i.e. \textbf{Motion\_1-8}).
They return `true' values if they detect  user movement. 
After a `true' value is detected, a `false' value is returned when motion is no longer detected near the sensor.
\newline
All motion data points are represented as `Motion\_X' in the `sensor\_name' column of the CSV files, where X ranges from 1 to 8.
Sensor values that are either [True, False] are recorded in the `value' column of the CSV files.
\item \textbf{Seat} - Each seat sensor is used to detect the occupancy of each particular seat and is assigned a unique number in its sensor name based on location (i.e. \textbf{Seat\_1-12}).
The seat sensor uses the pressure on the plate, thus, it returns `true' if force is applied to the plate on the seat (i.e. sitting action).
If a person gets up from the seat, the sensor returns `false'.
\newline
All data points from seat sensors are displayed as `Seat\_X' in the `sensor\_name' column of the CSV files, where X ranges from 1 to 12.
Their values are either [True, False] and are recorded in the `value' column of the CSV files.
\end{enumerate}

\subsubsection*{Actuator-driven data}
Actuators are devices that users can operate in a specific space, and actuator-driven sensors track their status.
DOO-RE includes data from three actuators: \textbf{AirCon}, \textbf{Light}, and \textbf{Projector}. 

\begin{enumerate}
\item \textbf{AirCon} - 
The sensor captures the power status of the air conditioner.
\textbf{Aircon\_1} and \textbf{Aircon\_2} are sensors that track the power status of air conditioners located in the front and back sides of the meeting room, respectively.
An AirCon data point occurs when an air conditioner is powered on and off. 
\newline
All \textbf{AirCon} data points are indicated as `Aircon\_X' in the `sensor\_name' column of the sensor data file, where X is 1 or 2.
The [On, Off] values of each data point are displayed in the `value' column of the CSV file.

\item\textbf{Light} - 
The sensor detects the light's power state and publishes a data point when the state changes.
Each light's name is associated with its respective location.
We refer to a light close to the left wall as \textbf{Light\_1}, a light in the middle as \textbf{Light\_2}, and a light close to the right wall as \textbf{Light\_3}.
When a light is turned on and off, a Light data point is recorded.
\newline
All \textbf{Light} data points are indicated as `Light\_X' in the `sensor\_name' column of the CSV files, where X is 1, 2, or 3. 
Their values are recorded as either [On, Off] in the `value' column of the CSV files.

\item \textbf{Projector} - 
The \textbf{Projector} sensor captures the power state of the projector. 
Projector data points occur when users turn the projector on and off.
\newline
All \textbf{Projector} data points are indicated as `Projector' in the `sensor\_name' column and their [On, Off] values are recorded in the `value' column of the CSV files.

\end{enumerate}

\begin{figure}[!t]
    \centering
    \includegraphics[width=\textwidth]{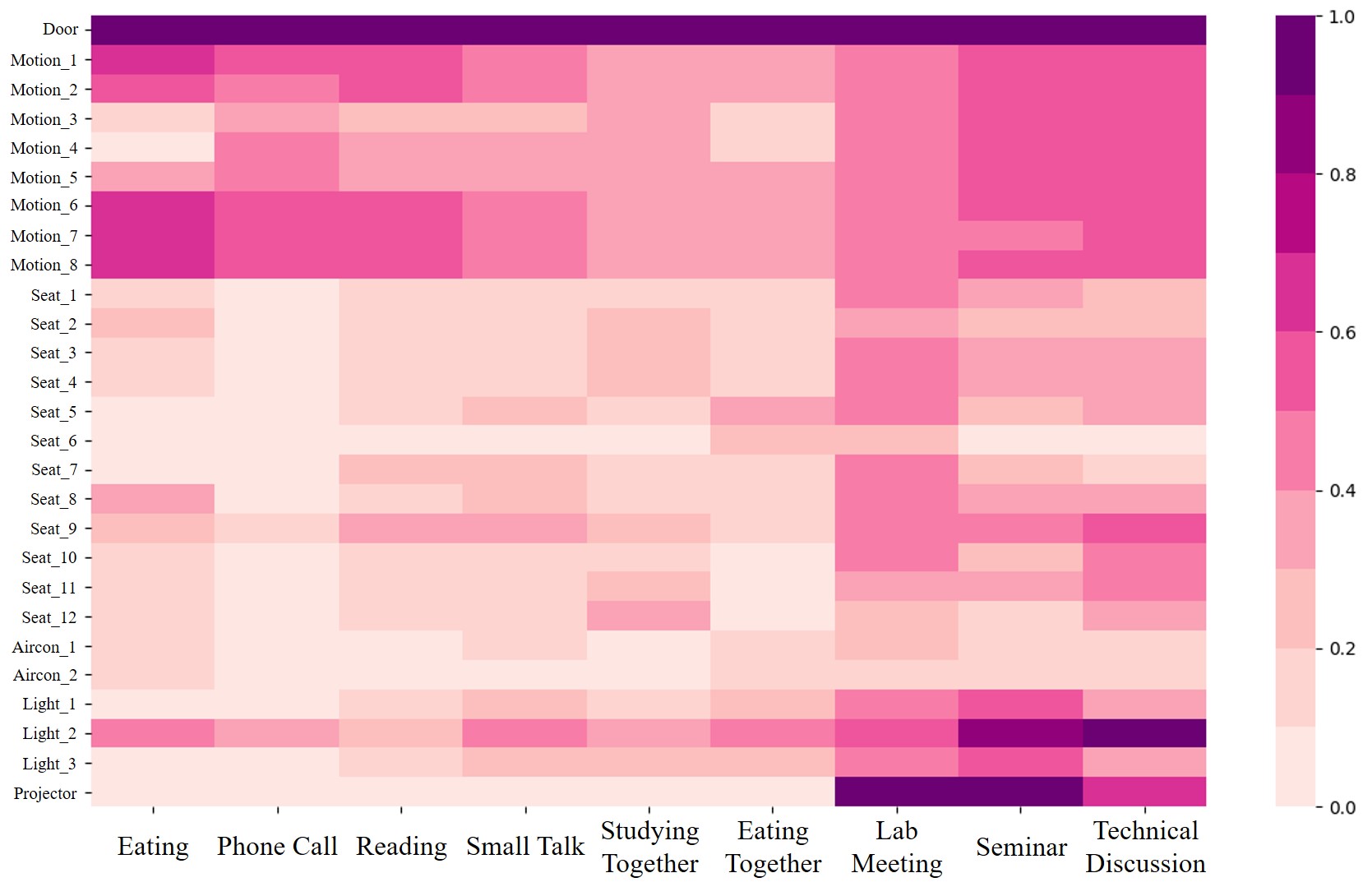}
    \caption{
    Visualization of whether each sensor occurs or not in an activity episode.
    The figure helps us roughly understand the characteristics of each activity.
    The darker the color, the more likely a certain sensor appears in an activity episode.
    For example, on average, `Projector' appear in all episodes of \textit{Lab meeting}, but rarely in all episodes of \textit{Eating together}.
    }
    \label{fig: sensor_activation}
\end{figure}

\begin{figure}[!t]
    \centering
    \includegraphics[width=\textwidth]{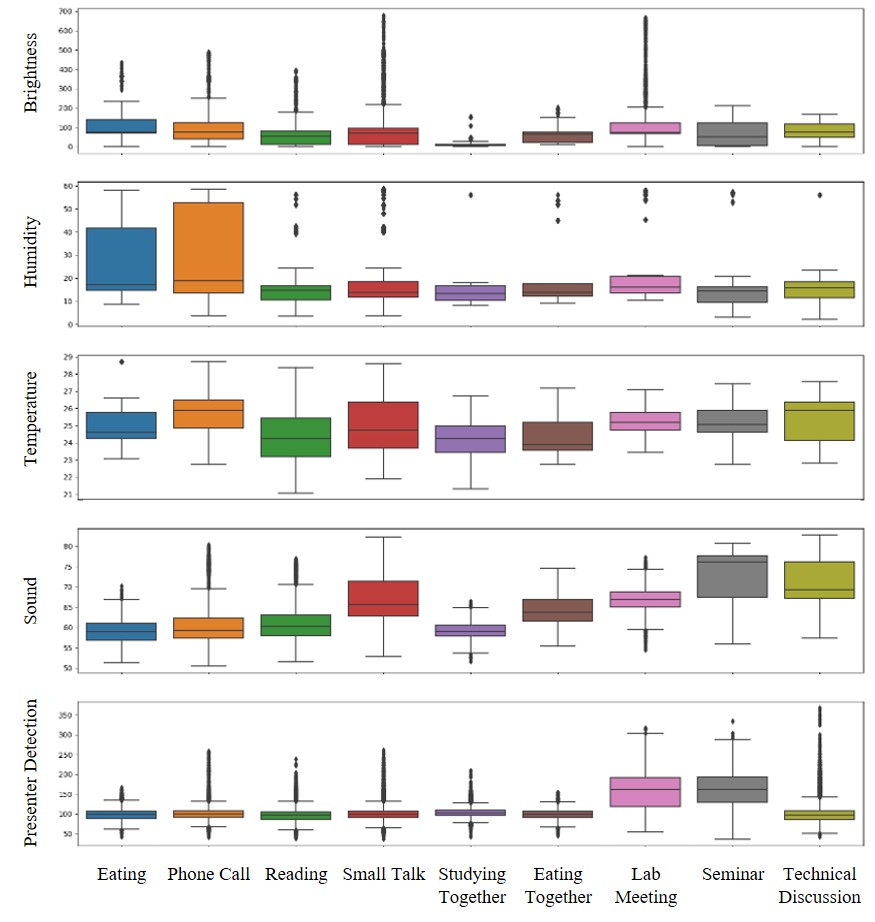}
    \caption{Value distributions of `Environmental-driven' sensors and the `Presenter Detection' sensor. Those sensors produce numerical values as states, thus, they are represented by numerical distributions.
    Each box plot represents the distribution of values across all episodes of each activity.}
    \label{fig: sensor_distribution}
\end{figure}

\begin{figure}[t]
    \centering
    \includegraphics[width=0.8\textwidth]{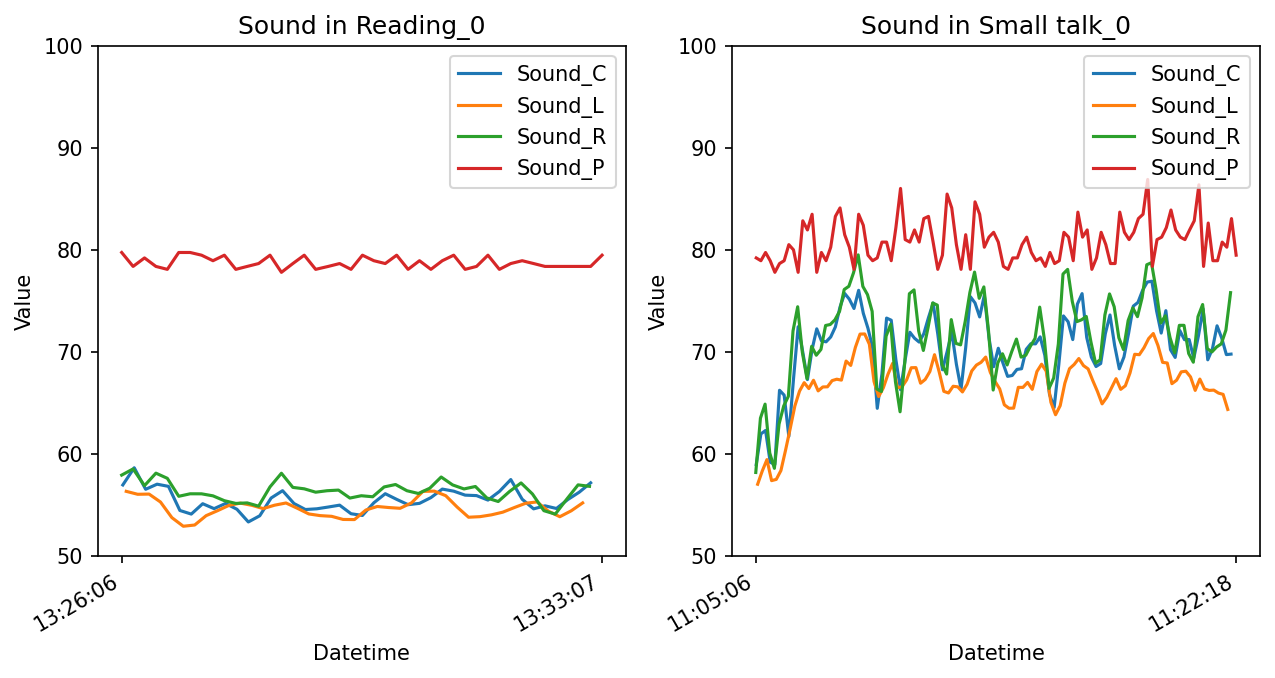}
    \caption{A graph of changes of `Sound' sensors' values in \textit{Reading\_0.csv} and \textit{Small talk\_0.csv} files.
    In the \textit{Reading\_0} activity, each `Sound' sensor generates stable and low values, but they generate data that fluctuates greatly between high and low values during \textit{Small talk\_0} activity where participants in the room talk to each other.
    }
    \label{fig: sensor_sound}
\end{figure}

\begin{figure}[!t]
    \centering
    \includegraphics[width=\textwidth]{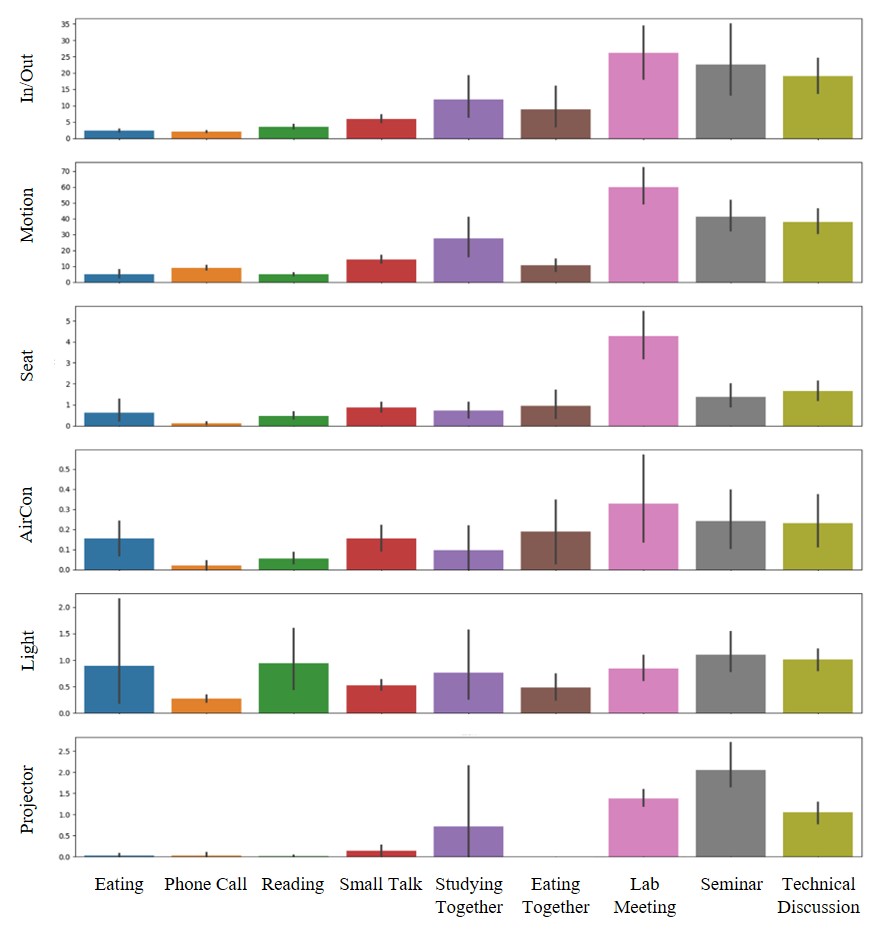}
    \caption{Frequency distributions per episode of `User-driven' and `Actuator-driven' sensors. Those sensors report states in the form of binary values, such as `True' or `False', which are better defined in terms of activation frequencies.
    Each box plot represents the frequency distribution of how many times each sensor is present per activity episode.}
    \label{fig: sensor_frequency}
\end{figure}

\section*{Technical Validation}
\subsection*{Annotation validation}
Activity labels are approximately 96\% consistent during cross-checks between annotators.
It implies that most episodes are perceived as agreed labels among them.
The remaining 4\% episodes, which have mismatched activity labels, are determined through discussion between annotators.
The reason why such vaguely labeled episodes come out is that we cannot record users' conversations in the meeting room or capture the status of personal objects such as laptops due to privacy issues.
For example, \textit{Small Talk} is often confused with \textit{Technical Discussion} as the annotators cannot know the conversation between users.
The labeling of \textit{Phone Call} and \textit{Reading} is also confusing because it is not known whether a user uses a phone to make a call or to read.
For these episodes, their final activity labels are determined by the majority votes of the annotators.

As shown in Table \ref{tab: activity information}, the number of episodes is disproportionate between activities due to the nature of the meeting room.
Activities that use the meeting room relatively lightly, such as \textit{Phone call}, \textit{Reading}, and \textit{Small talk}, occur more frequently than other activities.
The average duration and the average number of participants also depend on the activity type.
The single-user-based activities have shorter durations than the group-user-based activities as expected.
Among the group-user-based activities, except for \textit{Studying Together}, the average duration increases as the average number of participants increases.

Figure \ref{fig: sensor_activation} shows which sensors are activated on average for the activities of interest.
It expresses the number of episodes in which a particular sensor is involved out of the total number of episodes for a particular activity.
This can tell which sensors are important for understanding each activity.
`Environment-driven' sensors and the `Presenter Detection' sensor are not included in Figure \ref{fig: sensor_activation} because they occur in every activity.
In the case of group-user-based activities with a large number of participants (i.e. \textit{Lab Meeting}, \textit{Seminar}, \textit{Technical Discussion}), all sensors are more likely to be activated than other activities.
Especially, `Seat' sensors and `Actuator-driven' sensors help clarify the characteristic of these group-user-based activities.
The following section details the differences in characteristics between the activities.

\subsection*{Sensor distribution differences between activities}
Figure \ref{fig: sensor_distribution} shows box plots that describe value distributions of `Environmental-driven' sensors and the `Presenter Detection' sensor. 
These sensors generate numerical values as states, thus, they are represented by numerical distributions.
Figure \ref{fig: sensor_frequency} shows the average frequencies of the remaining sensors (Actuator-driven sensors and User-driven except for the `Presenter Detection' sensor).
They publish states as binary values such as `True' or `False' that are more appropriately described in terms of frequencies (i.e. number of occurrences).
Both figures show that sensor distributions and frequencies differ between activities, implying the superior ability of DOO-RE to represent these activities.
The sensor distributions and frequencies are consistent with what is thought of as common sense.

Figure \ref{fig: sensor_distribution} presents how the values of the `Environmental-driven' sensors and the `Presenter Detection' sensor vary depending on the activity.
`Brightness', `Humidity', and `Temperature' sensors generate similar values between activities as they are influenced by the external environment such as time of day or weather rather than user actions.
\textit{Phone Call} and \textit{Eating} show particularly large `Humidity' value range variations, presumably because these two activities can occur throughout the day compared to other activities.
Higher values of `Sound' sensors occur frequently in group-user-based activities since more people are talking than in single-user-based activities.
An example of this is the difference between `Sound' sensors' values in \textit{Reading} and \textit{Small Talk} in Figure \ref{fig: sensor_sound}.
Each `Sound' sensor generates stable and low values in the \textit{Reading} activity. 
However, during the \textit{Small talk}, the sensor fluctuates between high and low values as users in the room talk to each other.
The values of `Sound' are kept low in the`Studying together' activity since it is a quiet activity among group-user-based activities.
The `Presenter Detection' sensor, as expected, shows higher values in activities with presenters (i.e. \textit{Lab meeting} and \textit{Seminar}).

Figure \ref{fig: sensor_frequency} describes the number of occurrences per episode of `User-driven' and `Actuator-driven' sensors.
`User-driven' sensors have many activations per episode, while `Actuator-driven' sensors run only once or twice per episode.
The activation frequency per episode of each `User-driven' sensor (i.e. `In/Out', `Motion', and `Seat' sensors) is proportional to the number of participants.
This implies that `User-driven' sensors are useful for detecting changes in user states in the middle of an activity that occurs more frequently in large member-based activities.
In addition, we find that the difference in duration between activities affects the activation frequencies of the `User-driven' sensors.
For example, \textit{Studying together} generates more the `In / Out' and `Motion' sensors states than \textit{Eating together} since \textit{Studying together} lasts longer than \textit{Eating together}. 
The activation frequencies per episode of `Actuator-driven' sensors vary from activity to activity, depending on which actuators are used for each activity.
The activation frequencies of the `AirCon' sensor are  generally proportional to the number of participants in an activity.
The `AirCon' sensor frequencies per episode are indicated under 1 since the data collection period of the DOO-RE dataset is from autumn to winter, when air conditioners are not used much.
The `Light' sensor is roughly associated with the duration of the activities, and it appears in activities lasting more than a certain amount of time.
\textit{Phone Call} activates the `Aircon' or `Light' sensors much less than others since it is a simple and short activity where users enter the meeting room and make calls without manipulating any devices.
Activities using projectors, such as \textit{Lab meeting}, \textit{Seminar}, and \textit{Technical discussion}, primarily activate the `Projector' sensor, as expected.

Cross-interpretation of Figure \ref{fig: sensor_distribution} and Figure \ref{fig: sensor_frequency} provides a deeper understanding of a given activity.
In general, group user-based activities elicit higher values from `User-driven' and `Actuator-driven' sensors than single-user-based activities.
Among group-user-based activities, those that involve more people such as the \textit{Lab meeting}, \textit{Seminar}, and \textit{Technical Discussion} generate higher sensor values per episode throughout the sensors.
We explore that we can distinguish similar-looking activities by interpreting several types of sensors together.
Through `PodiumIR' and `Projector', we find how \textit{Seminar} and \textit{Lab meeting} are 
different from \textit{Technical Discussion}.
In \textit{Technical Discussion}, `Projector' statuses are similar to the other two activities, but the values of `PodiumIR' are noticeably smaller since there is usually no presenter in \textit{Technical Discussion} episodes.
As another example, \textit{Small talk} and \textit{Studying together} look similar, but they can be distinguished by the `Sound' level and the number of `Motion' activations.
This verifies us that by employing various sensors,  activities can be distinguished even in the meeting room that cannot be divided into physically independent sub-regions.

\subsection*{Data quality}
We investigate the data quality of ambient sensors in the DOO-RE for each activity.
User-driven and Actuator-driven sensor values may not appear in some episodes since their states are not published unless users interact with the corresponding objects or devices.
To differentiate between a sensor's missing value and a sensor's inactive state, we have to manually match video and sensor data to determine if that sensor actually worked.
We share the data quality results in \textbf{data\_quality.xls} file. 

In general, missing or inaccurate values are usually caused by problems such as sensor battery, network, physical sensor, or malfunctioning agents problems. 
For example, \textit{Digi XBee sensor /L/T/H}, which constructs \textbf{Brightness}, \textbf{Humidity}, and \textbf{Temperature} sensors, is a wireless device and their rapid battery consumption causes data to be missed.
In addition to the problem of missing or inaccurate sensor values, the sensors often record their values differently than we expected due to the nature of public space.
Each sensor may have missing or unexpected values for the following reasons: 

\begin{itemize}
  \item \textbf{Sound} sensors cannot distinguish between ambient noise and users' voices since they do not record human voices. For this reason, if sensors are activated by noise such as construction sounds, the corresponding episodes inevitably record higher decibels than other episodes.
  This means that sensor values may be recorded differently from the contexts we expect for \textbf{Sound} sensors.

  \item If the podium's position is changed by a presenter, the range of sensor values for \textbf{Presenter Detection} may vary.
  It can cause unexpected \textbf{Presenter Detection} sensor values.
  
  \item  An \textbf{In / Out} sensor measures its state by IR distance. 
  If a user accesses a meeting room door but does not enter the room, the sensor's value may be generated unexpectedly.
  In addition, if a user enters quickly from a certain angle, the sensor's value may be missing or incorrectly generated.
  
  \item If a person stands outside an angle that a \textbf{Motion} sensor cannot detect, the sensor can be missed the state. In addition, if a user passes too quickly, a \textbf{Motion} sensor may fail to detect the user's movement or generate fluctuated values.
  
  \item A \textbf{Seat} sensor generates its value from the sensitive pressure of a seat plate, producing `True' values simply by moving the corresponding seat, even if a user is not actually seated.
  In addition, a certain \textbf{Seat}  may not be correctly positioned in the location described in Figure \ref{fig: sensor_location} because a user may change the position of the \textbf{Seat}.
  These cases cause unexpected values of \textbf{Seat} sensors.
  
  \item  Due to university regulations, every air conditioner may be turned off regardless of the user's intention since they are centrally controlled.
  \textbf{Aircon} sensors may not be able to capture the `Off' value.
  
  \item For all actuators (i.e. \textbf{AirCon}, \textbf{Light} and \textbf{Projector}), if a user operates an actuator quickly several times, the corresponding sensor cannot properly track its status.
  For example, if a user turns a \textbf{Projector} `On' and `Off' more than three times within one minute, the `Off' command may disappear as the \textbf{Projector} needs time to turn `On'.
 
\end{itemize}

We continue to update possible causes of sensor value problems on the DOO-RE web page.

\section*{Usage Notes}

\subsection*{Potential applications}
DOO-RE is useful for assessing how robust a proposed recognition application is in the wild with group users present.
Previous studies \cite{group_HAR1, group_HAR2} using DOO-RE show that schemes that work well on DOO-RE also suit well on existing ambient sensor-based datasets.
The results show that DOO-RE has the potential to act as a helper to improve group activity recognition performance by providing multi-sensor perspectives of users' behavior.
In addition to those intelligence-building studies, a recent study \cite{online_HAR} shows that DOO-RE helps verify online recognition methods.
Sensor-based recognition techniques can be used in various applications ranging from healthcare to human-machine interaction \cite{HAR_application}, and DOO-RE, which targets the domain of public space that existing datasets do not cover, can play a major role in expanding the techniques.
In addition, DOO-RE can be applied to learn user preferences \cite{preference_learning} for optimized smart service provision.
While researches in this domain still focus on a single user, adding multiple user-based concepts such as group dynamics \cite{GroupDynamics} through DOO-RE-based group user behavior analysis can develop more scalable smart services.


\subsection*{Discussion about DOO-RE}
The various sensors we install in our meeting room testbed can be applied in any space in general. However, the current sensor and activity information of DOO-RE is specialized in our meeting room.
The generated data are centered on group-user-based activities, and in addition to these activities, it is necessary to explore more types of user activities in order to increase the scalability of the DOO-RE.
Exploring multi-user activities, where multiple users perform different activities at the same time, is good for the purpose.
Datasets for multi-user activities are absent, especially in public spaces.
Accordingly, we plan to extend the DOO-RE dataset to multi-user/multi-activities by installing various sensors in the university lounge.
An early experiment shows that activities in the meeting room and in the lounge are distinctly different.

We find it difficult to gather dozens of episodes of specific group user-based activities a year in reality since group user-based activities for the same purpose only occur once or twice a week.
However, the data size of the data set is important for use in the latest technologies such as deep learning.
We construct a sustainable data collection system and data will be updated regularly to increase the size of DOO-RE.
For data sizes that may still be insufficient, we plan to introduce various time series-based data augmentation techniques \cite{TS_DA} to develop a data augmentation method suitable for the DOO-RE.

As results from sensor data analysis, we find that external information such as seasonality or time of day can also affect sensor state changes. 
Therefore, including such information is expected to help in more accurate user activity recognition in future studies.



\section*{Code availability}
All codes for extracting collected data, supporting activity annotations, converting collected data to formatted data record files, and analyzing sensor data are based on Python 3.7 and various python libraries.
The codes are available on our lab's GitHub site (\url{github.com/cdsnlab/ScientificData}).
The DOO-RE dataset files and data collection results are inquired on the dataset's website (\url{doo-re.kaist.ac.kr}).
DOO-RE's sensor data files are recommended to open them with an Excel program since they are in CSV format.
Metadata files are easily browsed or edited using any editor as they are in text format.

We collect data from ambient sensors, actuators, and a camera through our MQTT-based IoT system named Lapras.
The system stores the collected raw data in the MongoDB database and extracts sensor data preprocessed with Python codes to construct DOO-RE.
Information and code for this collected system can also be requested through the DOO-RE site or the corresponding author.
We are welcome to respond if you need any help or information about DOO-RE, except for requesting privacy-sensitive information such as videos.

\bibliography{sample}


\section*{Acknowledgements} 

DOO-RE was supported by Institute of Information \& communications Technology Planning \& Evaluation(IITP) grant funded by the Korea government(MSIT) (No.2019-0-01126, Self-learning based Autonomic IoT Edge Computing).

\section*{Author contributions statement}
H.K. designed and performed data collection, organized and preprocessed collected datasets, visualized and performed technical validation, and wrote the entire manuscript.
G.K. constructed dataset record files, visualized the distribution of sensors, and performed the annotation.
T.L. and K.K. performed the annotation, and helped to construct the collected datasets.
D.L. supervised the overall dataset configuration, advised this project, and reviewed the manuscript.

\section*{Competing interests} 
The authors declare no competing interests.

\end{document}